\documentclass[11pt]{article}
\usepackage[utf8]{inputenc}
\usepackage{amsfonts,url,epsfig, amsmath, amsthm, bm}
\usepackage{authblk}
\usepackage{geometry}
\usepackage{sectsty}
\usepackage{booktabs}
\usepackage[normalem]{ulem}
\usepackage{subcaption}
\usepackage{algorithm, algpseudocode}
\usepackage[hidelinks]{hyperref}
\usepackage{todonotes}

\usepackage{color}
\newcommand{\tc}[1]{\textcolor{red}{[TC: #1]}}

\usepackage{url,breakurl}
\newcommand{\po}{ {\rm post}}
\newcommand{\E}[0]{\mathbb E}

\sectionfont{\sffamily\bfseries\upshape\large}
\subsectionfont{\sffamily\bfseries\upshape\normalsize}
\subsubsectionfont{\sffamily\mdseries\upshape\normalsize}
\makeatletter
\renewcommand\@seccntformat[1]{\csname the#1\endcsname.\quad}
\makeatother

\newtheorem{remark}{Remark}[section]
\newtheorem{prop}{Proposition}[section]

\usepackage{natbib}
\makeatletter
\def\@maketitle{%
  \begin{center}%
  \let \footnote \thanks
    {\large \@title \par}%
    {\normalsize
      \begin{tabular}[t]{c}%
        \@author
      \end{tabular}\par}%
    {\small \@date}%
  \end{center}%
}
\makeatother

\title{\bf Approximate posterior recalibration\footnote{We thank the U.S. National Science Foundation and U.S. Office of Naval Research for partial support of this research.}\vspace{.1in}}

\date{18 Mar 2026}

\author{Tiffany Cai\footnote{Department of Statistics, Columbia University, New York}, \ Philip Greengard\footnote{Center for Computational Neuroscience, Flatiron Institute, New York}, \ Ben Goodrich\footnote{Institute for Social and Economic Research and Policy, Columbia University, New York}, \ and Andrew Gelman\footnote{Department of Statistics and Department of Political Science, Columbia University, New York}}

\begin{document}

\maketitle

\begin{abstract}
Bayesian inference is often implemented using approximations, which can yield interval estimates that are too narrow, not fully capturing the uncertainty in the posterior distribution. We address the question of how to adjust these approximate posteriors so that they appropriately capture uncertainty.  
We introduce two methods that extend simulation-based calibration checking (SBC) to widen approximate posterior uncertainty intervals 
to aim for marginal calibration. We demonstrate these methods in several experimental settings, and we discuss the challenge of calibration using posterior inferences and the potential for posterior recalibration of hierarchical models.


\end{abstract}

\section{Introduction}
When we fit large, complicated models using approximate algorithms, there is concern about the calibration of uncertainty statements. We would like our inferences to recover true parameter values or population quantities, to the extent that they are identifiable from the data and model, and we would like our uncertainty statements to capture the errors in estimation.

Assume the following scenario of Bayesian computation.  
A true scalar parameter $\theta$ has been drawn by Nature from its prior distribution, $p(\theta)$.  
Data $y$ are then drawn from the likelihood  $p(y|\theta)$.  An analyst observes the data and knows the prior and data model and hence can write the posterior, $p(\theta|y)\propto p(\theta)p(y| \theta)$.  
By construction, Bayesian posterior intervals are calibrated,  attaining nominal coverage with respect to the prior. If we (a) draw parameters $\theta^l$ for $l=1,\ldots,L$ from the prior, (b) generate data $ y^l$ from $\theta^l$, and (c) construct posterior intervals for $ \theta_{\rm post}^{s}$ conditioned on $y^l$, then those intervals should contain the true parameter $\theta^l$ the correct portion of the time, averaged over draws of $\theta^l$ \citep{Cook06}, a procedure that has been formalized as simulation-based calibration checking (SBC) \citep{talts2020validating,Modrak2022SimulationBasedCC}. 

However, often the exact posterior cannot be obtained because of computational reasons, so that the analyst instead fits a more tractable model or uses an approximate computational procedure such as variational inference \citep{blei2017variational}. 
In contrast to intervals obtained from exact posteriors, intervals obtained from approximate posteriors are not generally expected to produce correct coverage with respect to the prior, even with a large number of simulations.
At a practical level, we would still like to use approximate posteriors, especially when exact posterior inference is too computationally costly. However, we can adjust approximate intervals with the aim of achieving posterior calibration. 

This paper makes two contributions. First, we propose recalibration methods based on simulation-based calibration checking for widening approximate posterior intervals so that they are calibrated with respect to the prior in Section~\ref{sec:methods}. Essentially, we pick a class of potential adjustments, and choose the adjustment that best calibrates posteriors according to SBC. We demonstrate these recalibration methods in several settings in Section~\ref{sec:experiments}. 

Second, we explore nuances in the problem of posterior recalibration, including the distribution with respect to which we would like to be calibrated. In particular, we consider the option of replacing step (a) above (drawing $\theta^l$
from the prior, $p(\theta)$), with draws of $\theta^l$
from $p(\theta| y)$ the posterior given
observed data (with other changes, including how the posterior draws $\theta^s_{\rm post}$ are calculated).
The advantage of using posterior draws is that the recalibration is focused on the zone of
parameter space that is of applied interest given the problem at hand. The disadvantage is that we
should no longer expect calibration, even if the approximate inference were exact---that is, 
where exact posteriors are generated in step (c) above. 
In Section~\ref{sec:normalnormal} we work out the details of simulation-based recalibration for a
one-dimensional normal-normal model, a simple setting where it is possible to obtain exact posterior draws and work out results analytically.  When using this exact computation, inferences averaging over the prior predictive
distribution are calibrated (up to Monte Carlo error), but inferences averaging over the posterior
predictive distribution are not, outside of extreme cases. Performing posterior predictive recalibration results in inferences that differ from the exact posterior by doing less pooling toward the prior.
In Section~\ref{posterior.hier} we discuss implications for simulation-based calibration and how 
posterior recalibration could still work well for hierarchical models, even while being so far off in a simple non-hierarchical setting. 

Our work is related to that of \cite{rodrigues2018}, who propose a method for recalibrating inferences constructed using the ABC procedure, \cite{yu2019assessment}, who propose a method of calibration checking and recalibration based on moments rather than quantiles, \cite{bon2022}, who propose an 
approach for optimizing a linear recalibration based on approximate simulations, and \citet{bls2}, who recalibrate posterior intervals by leveraging SBC performed on a different inference problem in which the prior in step (a) is replaced by the posterior. Our work is also related to \cite{lee2019calibration} and \cite{xing2019calibrated}, who estimate coverage of credible intervals obtained by an approximate posterior method, and also to \cite{xing2020distortion}, who estimates a ``distortion map'' from an approximate posterior to the true posterior, where the true posterior is approximated by ABC.

\section{Simulation-based calibration checking}
\label{sec:sbc}
In this section, we introduce notation and describe simulation-based calibration checking~\citep{Cook06,talts2020validating,Modrak2022SimulationBasedCC}, which we extend in our recalibration procedures. 
We use the notation $M$ for the model that we are trying to fit, which we also assume is the true model that generates the data.
We use the notation $\Theta\in\mathbb R^d$ for the vector parameters in the model, 
$\theta$ for a scalar component or function of $\Theta$ of interest (e.g. $g(\Theta)$ for some $g:\mathbb R^d\to \mathbb R$),
$y$ for the data, and $M^{\prime}$ for the approximate computation for the true model $M$. For simplicity, we let $g(\Theta):=\theta$, though all results extend immediately to more general scalar functions $g$. 
We are interested in the coverage of posterior intervals for $\theta$, averaging over the assumed joint distribution, $p(\Theta,y)$, and our concern is with the calibration of approximate posteriors under the true model $M$.

A natural way to check if the posteriors are correct is through simulation-based calibration checking as follows in Algorithm~\ref{alg:sbcc} (numbered so that our proposed algorithms, which extend Algorithm~\ref{alg:sbcc}, are numbered starting at 1): 

\begin{algorithm}[H]
\caption{\em Simulation-based calibration checking}\label{alg:sbcc}
 \textbf{Requires} Models $M,M'$, constants $L, S$, scalar component of interest $\theta$ in $\Theta$
\begin{algorithmic}[1]
\State Obtain $L$ independent draws 
$\Theta^1,\dots,\Theta^{L}$ of $\Theta$ from its prior distribution in model $M$. 
\label{step1}
\For{$l = 1,\dots,L$, in parallel}
        \State Draw 
        $y^l\sim p(y | \Theta^l)$
        from the data model in $M$.
        \State Fit $M^{\prime}$ to $ y^l$ to obtain $S$ independent approximate posterior draws of  
        $\theta_{\rm post}^{l,s},\ s=1,\dots,S$.
        \State Determine the quantile $q^l$ of $\theta^l$ in this distribution, that is, $q^l=\frac{1}{S}\sum_{s=1}^{S} 1_{\theta^l > \theta_{\rm post}^{l,s}}$
      \EndFor
        \State If the model is fit exactly (that is, if $M^{\prime}\equiv M$) and the simulation draws are independent, then $q^1,\dots,q^L$ should be independent random draws from the uniform distribution on $(0,1)$. 
\State Compare $q^1,\dots,q^L$ to the uniform distribution on $(0,1)$.  A discrepancy indicates miscalibration.
\end{algorithmic}
\end{algorithm}
Simulation-based calibration checking~\citep{talts2020validating,Modrak2022SimulationBasedCC} relies on the well-known observation that the data-averaged posterior and the prior are self-consistent~\citep{Cook06}: 
{consider draws $\Theta^l$ from the prior $p(\Theta)$, $\theta^l\sim p(\theta)$ the scalar component of interest $\theta$ from $\Theta^l$, and $y^l\sim p(y|\Theta^l)$.}
Then
by construction
$( \theta^l, y^l) \sim p(\theta, y)
$
so that
\begin{align}
\label{eq:identity2}
\theta^l 
\sim p(\theta|   y^l).
\end{align}
The parameter $\theta^l$, which was drawn from $p(\theta)$,
can also be thought of as being drawn from the posterior
$p(\theta| y^l).$ 
Checking calibration is ultimately verifying that $ \theta^l$
appears to be sampled from the posterior $p(\theta|  y^l)$,
which can be done by comparing $ \theta^l$
with independently drawn posterior samples from the posterior, 
$\theta_{\rm post}^{l,s}\sim p(\theta| y^l)$. 
From this we obtain the condition used in simulation-based calibration checking:
\begin{align}\label{110}
p(\theta)=\int \!\!\!\int p(\theta^l) \, p(  y^l| \theta^l) \, p(\theta| y^l) \, dy^l\, d \theta^l ,
\end{align}
{where $p(y^l|\theta^l)$ is the likelihood obtained by integrating out the remaining components of 
$\Theta$, denoted $\Theta_{-\theta}$, under the prior, $p(y^l|\theta^l) = \int\! p(y^l|\Theta^l)p(\Theta_{-\theta}^l|\theta^l) d\Theta_{-\theta}^l$.}
In this paper, we go beyond the above simulation-based calibration checking procedure. We don't want to just identify problems with calibration; we would also like to obtain roughly calibrated intervals. 

\section{Proposed procedures for approximate posterior calibration (APC)}
\label{sec:methods}
The central idea is that because the exact posterior satisfies data-averaged self-consistency 
\eqref{110},
the exact posterior satisfies various specific self-consistency properties. 
Now suppose we have a way to obtain approximate posteriors. We can choose a specific self-consistency property and learn an adjustment so that these approximate
posteriors satisfy this specific self-consistency property, too. 
Then we can apply the adjustment we learned in the previous step 
on an approximate posterior for a specific dataset $y_D$ that we care about. We call this general procedure \emph{approximate posterior calibration} (APC). The general algorithm is as follows:
\begin{algorithm}[H]
\caption{\em Approximate posterior calibration (general)}\label{alg:adjust}
 \textbf{Requires} Models $M,M'$, constants $L, S$, {scalar component of interest $\theta$ in $\Theta$}, data $y_D$, specific self-consistency property
\begin{algorithmic}[1]
\State Obtain $L$ independent draws {$\Theta^1,\dots,\Theta^L$ of $\Theta$} from its prior distribution in model $M$. \label{step1_2}
\For{$l = 1,\dots,L$, in parallel}
        \State Draw {$y^l\sim p(y | \Theta^l)$} from the data model in $M$.
        \State Fit $M^{\prime}$ to $y^l$ to obtain $S$ independent approximate posterior simulations, {$\Theta^{l,s}_{\rm post},\ s=1,\dots,S$.}
        \State {Restrict $\Theta^{l,s}_{\rm post},\ s=1,\dots,S$ to the scalar component of interest to obtain  $\theta^{l,s}_{\rm post},\ s=1,\dots,S$.}
\EndFor
\State Calculate the adjustment needed to satisfy the specific self-consistency property.
\State Fit $M'$ to $y_D$ to obtain an approximate posterior.
\State Apply the adjustment from Step 7 to the approximate posterior from Step 8.
\end{algorithmic}
\end{algorithm}
In the above, 
we have not specified what ``Calculate the adjustment needed to satisfy the specific property'' is,
as this step can vary with the specific property desired. 
Thus from the above we have a class of procedures to adjust 
a posterior approximation method, each corresponding to a way of obtaining
an appropriate adjustment for a posterior, which consists of
\begin{enumerate}
    \item A class of adjustments, and 
    \item A way to choose an adjustment from that class of adjustments. 
\end{enumerate}
For the sake of simplicity, in the following methods that we will propose, our class of adjustments will be just a fixed width scaling
$k$ for posteriors, where the scaling is around the mean, so that if the samples from the approximate posterior are $\theta^{l,s}_{\rm post}$ for $s=1,\ldots,S$, the 
corresponding sample from the adjusted posterior is
\begin{align}
k (\theta^{l,s}_{\rm post} - \mu^{l}_{\rm post}) + \mu^{l}_{\rm post},
\end{align}
where $\mu^{l}_{\rm post}$ is the mean of $\theta^{l,s}_{\rm post}$ for $s=1,\ldots,S$, and the scaling factor $k$ is chosen by the method.  Our approach could be generalized to the shift-and-scale adjustments considered by \cite{bon2022}. 

Now we discuss two simple ways of choosing adjustments. 
One can easily think of variations, but we stick to these for simplicity. 
\subsection{Nominal coverage method}
\label{sec:nominal_coverage_method}
The objective is for $(1-\alpha)$ intervals to achieve nominal coverage. 
From \eqref{eq:identity2}, the quantiles $q^l:=p(\theta^l\geq \theta_{\rm post}^{l})$, 
should be uniform across replications indexed by $l$
if $\theta^{l}_{\rm post}$ 
is the true posterior distribution of $\theta$ 
given $y^l$. Thus, $(1-\alpha)$ of the time, $ \theta^l$
should be contained within the interval from the $(\alpha/2)$-quantile to the $(1-\alpha/2)$-quantile of $\theta_{\rm post}^l$. 
In other words, the property we aim to satisfy is
$$q(1-\alpha/2)=1-\alpha/2 \quad \text{ and } \quad 
q(\alpha/2)=\alpha/2.$$
If we only have samples $\theta_{\rm post}^{l,s}$ for $s=1,\ldots,S$ that are from an approximate posterior, and if we believe that only the posterior variance requires correction,
then we can scale the posterior samples $\theta_{\rm post}^{l,s}$ by some $\sigma$ around its mean 
where the $\sigma$ is chosen to minimize the following objective:
$$(q(1-\alpha/2)-q(\alpha/2)) - (1-\alpha) )^2.$$ 
One way to find the best adjustment 
is to do a grid search over potential adjustment values $\sigma$
to find the minimum value of the objective above. 
\subsection{Method using $z$-scores}
\label{sec:zscore_method}

{
As before, 
consider  draws $\Theta^l\sim p(\Theta)$, $\theta^l\sim p(\theta)$ the scalar component of interest $\theta$ from draws $\Theta^l$, and $y^l\sim p(y|\Theta^l)$.
If $\theta_{\rm post}^{l,s} \sim p(\theta|  y^l)$ are draws from the true posterior, we can think of $ \theta^l$ as being drawn from $p(\theta |  y^l)$, from which we have draws $\theta_{\rm post}^{l,s}$. 
Let $\mu_{\rm post}^l$ be the mean of $\theta_{\rm post}^{l,s}$ across $s=1,\ldots,S$, and let $\sigma^l_{\rm post}$ be the standard deviation of $\theta_{\rm post}^{l,s}$ across $s=1,\ldots,S$. 
Then we can consider the $z$-score
for $\theta^l$, $z^l:= (\theta^l-\mu_{\rm post}^l)/\sigma^l_{\rm post}$. Across $l=1,\ldots,L$, under self-consistency, the $z$-scores $z^l$ should have mean 0 and variance 1. Thus, the
property we aim to satisfy is that  
$z^l\sim(0,1)$, i.e. 
\begin{equation}
\label{eq:z_score_condition}
\mbox{Self-consistency: } \ 
\E\left(\frac{\theta^l-\mu_{\rm post}^l}{\sigma^l_{\rm post}}\right) = 0 \; \mbox{ and } \;\mbox{sd}\left(\frac{\theta^l-\mu_{\rm post}^l}{\sigma^l_{\rm post}}\right) = 1,
\end{equation}
where the expectation and standard deviation are over draws $l=1,\ldots,L$. }

{For some posterior approximation methods, if we believe that only the posterior variance requires correction, it can be reasonable to focus on just the second property. Thus, 
a way to satisfy (\ref{eq:z_score_condition}) while assuming only the posterior variance requires correction
is to consider an adjusted version of the posterior sample,
\begin{align}
\label{eq:z_score_adjustment_scale}
    \theta_\po^{{\rm adj},l,s}:=\mu_{\rm post}^l + s_z(\theta^{l,s}_{\rm post} - \mu_{\rm post}^l), 
\end{align}
where $s_z$ is the standard deviation of $z^l$ across $l=1,\ldots,L$. 
By construction, if we assume the first self-consistency property in \eqref{eq:z_score_adjustment_scale} holds, then this adjusted posterior in \eqref{eq:z_score_adjustment_scale} satisfies the second self-consistency property in \eqref{eq:z_score_condition}; see Remark~\ref{rmk:zscore}.

{\begin{remark}[Adjusted posterior satisfies self-consistency property in \eqref{eq:z_score_condition}]\label{rmk:zscore}
We show that 
\begin{align}
    \theta_\po^{{\rm adj},l,s}:=\mu_{\rm post}^l + s_z(\theta^{l,s}_{\rm post} - \mu_{\rm post}^l) + \overline z \sigma^l_{\rm post},
\end{align}
satisfies the self-consistency property in \eqref{eq:z_score_condition}, where $\overline z$ and $s_z$ are the mean and standard deviation of $z^l$ across $l=1,\ldots,L$, respectively. 
Compared to \eqref{eq:z_score_adjustment_scale}, here we do not assume only the width of the posterior requires correction; we choose a simpler correction without the constant shift above in \eqref{eq:z_score_adjustment_scale}. 
Write $\mu^l_{\rm post}:=\E(\theta^{l,s}_{\rm post})$ and $\sigma^l_{\rm post}:={\rm sd}(\theta^{l,s}_{\rm post})$, the mean and standard deviation of posterior samples indexed by $s=1,\ldots,S$, respectively. Also write $z^l:=\frac{\theta^l-\mu^l_{\rm post}}{\sigma^l_{\rm post}}$. 
Consider adjusted posterior samples,
$$\tilde \theta^{l,s}_\po := k^l(\theta^{l,s}_\po - \mu^l_\po)+\mu^l_\po+c^l,$$ 
for some $k^l,c^l$. Then write 
$\tilde \mu^l_\po := \E(\tilde \theta^{l,s}_{\rm post})=\mu^l_\po+c^l$ and 
$\tilde \sigma^l_\po := {\rm sd}(\tilde \theta^{l,s}_{\rm post})=k^l \sigma^l_\po$ with expectations and standard deviations taken over $s=1,\ldots,S$. 
Then the $z$-scores for adjusted posteriors are,
\begin{align}
\label{eq:z_score_adjust_both}
    \tilde z^l:=\frac{\theta^l-\tilde \mu^l_{\rm post}}{\tilde \sigma^l_\po} = \frac{\theta^l-\mu_\po^l - c^l}{k^l \sigma^l_\po}=\frac{1}{k^l}\left(z^l-\frac{c^l}{\sigma^l_{\rm post}}\right).
\end{align} 
In order to satisfy the conditions in \eqref{eq:z_score_condition}, that
$\E(\tilde z^l)=0$ and ${\rm sd}(\tilde z^l)=1$ (with expectations taken over $l=1,\ldots,L$), 
we set $k^l=s_z$ 
and $c^l=\overline z \sigma^l_\po$, with both expectations and standard deviations taken across $z^l$.
\end{remark}}
\subsection{Comparison}
In contrast to the nominal coverage method, the $z$-score method does not directly aim for nominal coverage for specific $(1-\alpha)$ intervals;
there is only one scaling factor $s_z$ for all intervals. This can be a disadvantage if the posterior width scaling \emph{should} vary by $\alpha$. 
However, if the optimal posterior width scaling does not vary by $\alpha$, the $z$-score method has the advantage of being comparatively more stable to estimate than the nominal coverage method, especially for $\alpha$ that are close to 0 or 1. 
Unlike the nominal coverage method, the $z$-score method can be used to select $\sigma$ without having to search for the best value. 
Empirically, the two methods give similar values for $\sigma$ in our experiments in the following sections. 






\section{Calibration experiments}
\label{sec:experiments}
We demonstrate and evaluate our procedure by applying it 
to several examples that we already understand well, where we know the ground truth posterior and can use it to evaluate our procedure. 


\subsection{A simple example: a one-parameter Gaussian model}
\label{sec:simple}
To provide a concrete example, we apply the proposed APC methods to the following model:
\begin{equation}\label{simple_model}
\begin{aligned}
    \theta & \sim \text{normal}(0, 1) \\
    y|  \theta & \sim \text{normal}(\theta, 1).
\end{aligned}
\end{equation}
For simplicity, we let $\Theta=\theta$.
With this simple model we have the benefit of a formula for the posterior,
\begin{align}
\theta \sim \text{normal}\,\bigg( \frac{1}{N+1} \sum_{i=1}^N y_i,\, \frac{1}{\sqrt{N+1}}\bigg).
\end{align}
The primary purpose of our method is to calibrate confidence intervals for functions of the posterior when 
the posterior is difficult to sample from and only approximate inference methods 
are available. In the above model we're far from that regime---we can easily sample from the posterior using standard Hamiltonian Monte Carlo (HMC) methods, for example, or even with independent draws from the closed-form posterior.  We use the above model primarily to build intuition. 

We now demonstrate the calibration procedure applied to this 
simple model. We artificially introduce a sampling error by narrowing the posterior standard deviation by a scaling factor of $3$. 
Without knowing the ground truth, we can tell that the approximate posterior is not calibrated, by using quantiles from simulation-based calibration checking (SBC)~\citep{talts2020validating}, which we have also repeated here in Algorithm~\ref{alg:sbcc}. 
\begin{figure}
\centering
\begin{subfigure}[t]{0.45\textwidth}
	\centering
	\includegraphics[width=0.9 \textwidth]{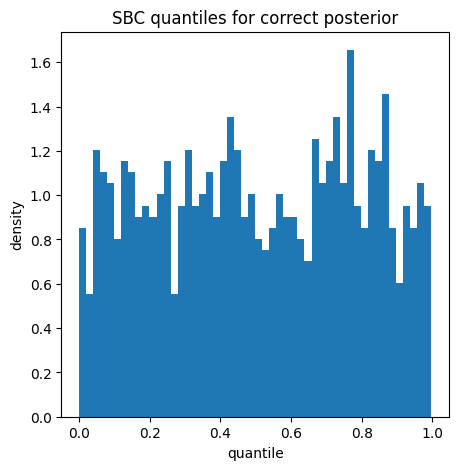}
	\caption{\em SBC quantiles for true posterior}
	\label{fig:schools_true_quantiles}
\end{subfigure}
\begin{subfigure}[t]{0.45\textwidth}
	\centering
	\includegraphics[width=0.9 \textwidth]{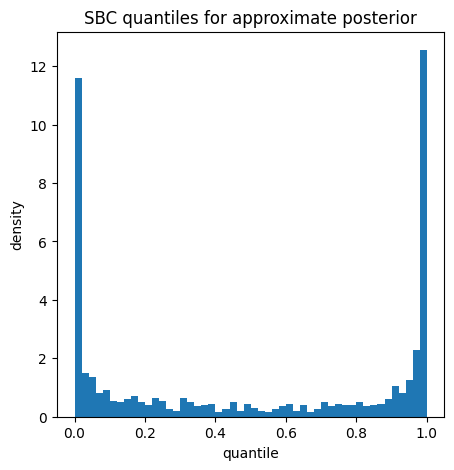}
	\caption{\em SBC quantiles for  posterior with sampling error}
	\label{fig:schools_approx_quantiles}
\end{subfigure}
\caption{\em SBC quantiles for model \eqref{simple_model} when sampling from the exact posterior (left) and from posterior with sampling error (right)}
\label{fig:simple_compare_calibration}
\end{figure}
Then, to calibrate $(1-\alpha)$ posterior intervals for $\alpha\in\{0.05,0.1,0.2,0.5\}$, we find scale adjustments using the nominal coverage method (Section~\ref{sec:nominal_coverage_method})
using a grid search, for scaling values $2.00,2.01,\ldots,5.00$. Indeed, using this method, we nearly recover the true posterior intervals for each value of $\alpha$ in Table~\ref{tab:simple_scaling_nominal},
as the scaling value obtained is usually close to 3. 
After calibration, the adjusted posteriors attain approximately 
nominal coverage (Table~\ref{tab:simple_scaling_nominal}). 
\begin{table}
    \centering
\begin{tabular}{ccc}
$1-\alpha$ & posterior width scaling $\sigma$ & adjusted coverage \\\hline
0.95 & 3.10 & 0.952 \\
0.90 & 2.98 & 0.896 \\
0.80 & 3.07 & 0.795 \\
0.50 & 3.17 & 0.491 
\end{tabular}
\caption{\em Posterior scaling of $(1-\alpha)$-level posterior intervals for the simple model \eqref{simple_model} using the nominal coverage method.}
\label{tab:simple_scaling_nominal}
\end{table}
We also find posterior scaling adjustments using the $z$-score method (Section~\ref{sec:zscore_method}), which gives
$\sigma=3.083$, which is also close to the scaling to recover the true posterior,
and which is more computationally efficient than the nominal coverage method as it does not require grid search. 
The adjusted posteriors attain almost nominal coverage by this method as well; see Table~\ref{tab:simple_scaling_z}. 
\begin{table}
    \centering
\begin{tabular}{ccc}
$1-\alpha$ & posterior width scaling $\sigma$ & adjusted coverage \\\hline
0.95 & 3.08 & 0.951 \\
0.90 & 3.08 & 0.909 \\
0.80 & 3.08 & 0.796 \\
0.50 & 3.08 & 0.475
\end{tabular}
\caption{\em Posterior scaling of $(1-\alpha)$-level posterior intervals for the simple model \eqref{simple_model} using the $z$-score method.}
\label{tab:simple_scaling_z}
\end{table}

\subsection{Hierarchical model (8 schools) and ADVI}
\label{sec:8schools}
We use the hierarchical modeling example for the 8 schools model, as defined and discussed in Chapter 5 of \cite{gelmanbda}. 

Here, we compute the posterior distribution using automatic differentiated variational inference (ADVI) {\citep{kucukelbir2017automatic}}, a fast approximate method.
We use ADVI to obtain approximate posteriors for the 8 schools example, and then we use our methods to adjust the approximate posteriors. 
We focus on the parameter $\mu$, the overall mean of the effect size across all schools. {Using notation from  Algorithm~\ref{alg:adjust}, $\Theta$ consists of all latent parameters in the 8 schools model, and we let the scalar parameter of interest $\theta$ be $\mu$.} 
Later, to sanity check, we compare the unadjusted and adjusted ADVI posteriors to a posterior computed using 
Hamiltonian Monte Carlo (HMC), 
which we consider to be ground truth
after checking simulation-based calibration on that posterior~\citep{Modrak2022SimulationBasedCC,talts2020validating}. Both ADVI and HMC are used via their respective implementations in Stan \citep{carpenter2017stan,cmdstanpy}.

We use ADVI on a centered parameterization and HMC on a non-centered parameterization 
for the 8 schools example, with parameterizations as discussed in \cite{betancourt2015hamiltonian}, as it is easier to produce a correct posterior using
a non-centered parameterization than a centered parameterization, for the purposes of 
demonstrating our posterior adjustment method. 
\begin{figure}
\centering
\begin{subfigure}[t]{0.45\textwidth}
	\centering
	\includegraphics[width=0.9 \textwidth]{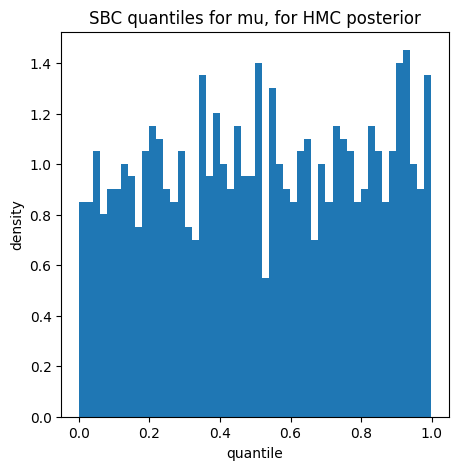}
	\caption{SBC quantiles with HMC}
	\label{fig:schools_hmc_quantiles}
\end{subfigure}
\begin{subfigure}[t]{0.45\textwidth}
	\centering
	\includegraphics[width=0.9 \textwidth]{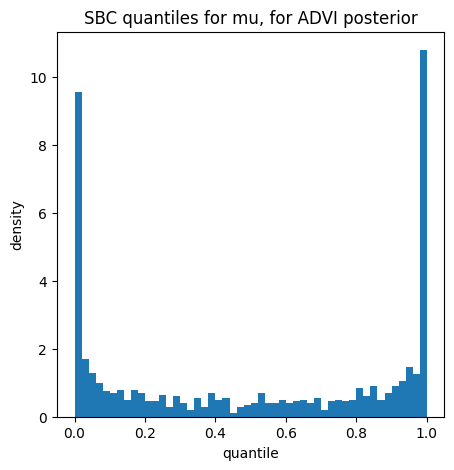}
	\caption{SBC quantiles with ADVI}
	\label{fig:schools_advi_quantiles}
\end{subfigure}
\caption{\em SBC quantiles for the 8 schools model (Section~\ref{sec:8schools}) when sampling with HMC (left) and ADVI (right). The HMC posterior is calibrated, while the ADVI posterior is not.}
\label{fig:schools_compare_calibration}
\end{figure}


Although the calibration process requires repeated computation, the posterior inferences
can all be run in parallel, so that the total run  time can be comparable to that of 
just one posterior inference computation. Furthermore, the calibration adjustment
can be computed for multiple $(1-\alpha)$ intervals after approximate posteriors are computed,
and also, once an adjustment is computed for $(1-\alpha)$ intervals, the adjustment can be directly applied 
to multiple datasets $y$.

We sample $L=1000$ draws of $\Theta^l$ (and also $\theta^l$), and for each $\Theta^l$, we generate $S=1000$ 
draws of  $\theta^{l,s}_{\rm post}$. 
From Figure~\ref{fig:schools_compare_calibration} where we calculated SBC quantiles~\citep{talts2020validating}, we see that the approximate posterior from ADVI is not calibrated.
Then, the appropriate scalings we find using grid search over $\sigma$ are displayed in Table~\ref{tab:8_schools_scaling_nominal}. 
\begin{table}
    \centering
\begin{tabular}{ccc}
$1-\alpha$ & posterior width scaling $\sigma$ & adjusted coverage \\\hline
  0.95   & 2.41 & 0.949 \\
  0.90   & 2.41 & 0.987 \\
  0.80   & 2.56 & 0.802 \\
  0.50   & 2.46 & 0.498 
\end{tabular}
\caption{\em Posterior scaling of $(1-\alpha)$-level posterior intervals from ADVI for the 8 schools example using the nominal coverage method.}
\label{tab:8_schools_scaling_nominal}
\end{table}
We compare to the $z$-score method in Table~\ref{tab:8_schools_scaling_z}. 
\begin{table}
    \centering
\begin{tabular}{ccc}
$1-\alpha$ & posterior width scaling $\sigma$ & adjusted coverage \\\hline
0.95 & 2.48 & 0.958 \\
0.90 & 2.48 & 0.909 \\
0.80 & 2.48 & 0.788 \\
0.50 & 2.48 & 0.505 
\end{tabular}
\caption{\em Posterior scaling of $(1-\alpha)$-level posterior intervals from ADVI for the 8 schools example using the $z$-score method.}
\label{tab:8_schools_scaling_z}
\end{table}
We see that the posterior width scaling $\sigma$ is similar under the nominal coverage method
and the $z$-score method. Accordingly, the 
$(1-\alpha)$ intervals 
for the corresponding adjusted posteriors also have similar coverage. 

\begin{figure}
\centering
\begin{subfigure}[t]{0.45\textwidth}
	\centering
	\includegraphics[width=0.9 \textwidth]{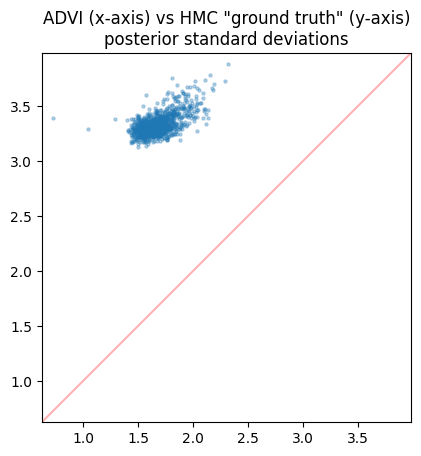}
	\caption{Posterior standard deviation\\for ADVI vs.\ HMC (``ground truth'')}
	\label{fig:schools_compare_sd}
\end{subfigure}
\begin{subfigure}[t]{0.45\textwidth}
	\centering
	\includegraphics[width=0.9 \textwidth]{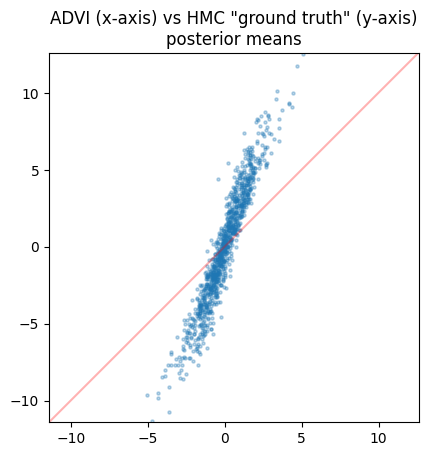}
	\caption{Posterior mean\\for ADVI vs.\ HMC ("ground truth")}
	\label{fig:6b}
\end{subfigure}
\caption{\em Posterior mean and standard deviation for ADVI vs.\ HMC.  Each dot represents a different replication from the prior predictive distribution {(i.e. a different $l\in\{1,\ldots,L\}$ in  Algorithm~\ref{alg:adjust})}.}
\label{fig:schools_compare_post}
\end{figure}

Figure~\ref{fig:schools_compare_post} contextualizes the posterior adjustments we found by comparing the ADVI (approximate) and HMC (``ground truth'') posteriors.  The ADVI posteriors are about twice as wide. This is smaller than the scalings
in Table~\ref{tab:8_schools_scaling_nominal} and~\ref{tab:8_schools_scaling_z}.
Nevertheless, the larger scaling we found is sensible because in Figure~\ref{fig:schools_compare_post} 
the ADVI posteriors are shifted (rather than only scaled) from the HMC posteriors, where the shift varies by $\theta^l$ 
and appears to be centered around 0. 
Because of this shift, widening ADVI posteriors by only a factor of 2 
isn't enough for posterior intervals to contain the true draw $\theta^l$ the correct proportion
of the time. Thus, in the absence of a way to adequately correct for the variation in shift, the 
next best thing to do is to widen posterior intervals by the scaling factors that we found.

\if 1=0
\section{Hierarchical models (8 schools): parameterization}
For the 8 schools prpoblem, Hamiltonian Monte Carlo (HMC) performs well under the so-called uncentered parameterization but hasd problems under the centered parameterization). \tc{describe and cite}
Here, we adjust a posterior that was obtained using the poorly parameterized model,
and 
we evaluate our adjustments using the well-parameterized model as ground truth. 
This example shows that adjustment can be useful not only for approximate posterior
methods, but also for MCMC with poorly parameterized models. 

\tc{we didn't actually get anything sensible here. Either best shift/scale was 1, or the best shift/scale was not 1 and kind of wrong but the difference in scores was very small. I wonder if it's because these are already pretty close to the truth}

\section{Linear regression}
For the linear regression example, we use the hierarchical model described
in (fastNoNo paper). This model has the advantage that we have fast and 
accurate algorithms for computing posterior moments. However, it is 
still a simplification of the model we would prefer to fit, which would 
involve fitting several group-level variances. 
We will use a model of the form 
\begin{align}
    y & \sim \text{normal}(X_1 \beta_1 + X_2 \beta_2, \sigma_y) \\
    \beta_1 & \sim \text{normal}(0, \sigma_1) \\
    \beta_{2, i} & \sim \text{normal}(0, \sigma_{2, i}) \\
    \sigma_y & \sim \text{normal}^+(0, 1) \\
    \sigma_1 & \sim \text{normal}^+(0, 1) \\
\end{align}
where $\sigma_{2, i}$ is the fixed hyperprior scale parameter on the
regression coefficient $\beta_{2, i}$.
\fi



\section{Posterior recalibration}\label{sec:normalnormal}

An intriguing alternative is to replace the prior distribution with the approximate posterior distribution 
in Step~\ref{step1_2} in Algorithm~\ref{alg:adjust}, so that averages over the prior are replaced by averages over the approximate posterior, and no other changes are made to the algorithm.  This has the benefit of focusing the recalibration on the zone of parameter space that is consistent with the data. This may be desirable, for example, if the prior has a lot of mass on parameter values that do not require recalibration, but the data are consistent with parameter values that do require substantial recalibration.

Posterior predictive recalibration cannot work in general, as we shall demonstrate with a simple normal-normal example, working out the details of prior and posterior recalibration. Applying the recalibration procedure to the posterior predictive distribution has the effect of reducing the pooling toward the prior. That said, we think posterior recalibration has potential when applied to hierarchical models, by making use of the internal replication of intermediate parameters. 

\subsection{Understanding posterior recalibration for a one-parameter normal model}

We can examine the differences between prior and posterior recalibration by considering a simple non-hierarchical model with only one parameter.  In this scenario, the model contains no internal replication, and there is no reason to expect posterior recalibration, even approximately.

We consider the following simple normal-normal model with one scalar parameter: 
\begin{eqnarray}
\nonumber  \theta &\sim& \mbox{normal}(0, 1) \\
  \label{model.0}  y|\theta &\sim& \mbox{normal}(\theta, \sigma),
\end{eqnarray}
with $\sigma$ assumed known.
This setup is more general than it might look, as $y$ can be taken to be not just one observation but rather as any unbiased estimate such as a sample mean or regression coefficient.  A proper prior is necessary here, as otherwise it would not be possible to perform prior predictive evaluation.

For our example, it is trivial to draw directly from the posterior, so we can evaluate the properties of calibration checking without needing to define an approximate posterior. 

\subsection{Prior calibration checking}



First we check that prior calibration checking works here, as we know it should in general, from \cite{Cook06}.  With prior predictive checking, the replications $l=1,\dots,L$ have the following distribution:
\begin{eqnarray}
 \label{model_1a} \theta^l &\sim& \mbox{normal}(0, 1) \\
  \label{model_1b}  y^l| \theta^l &\sim& \mbox{normal}(\theta^l, \sigma)\\
  \label{model_1c}  \theta_{\rm post}^{l,s}| y^l &\sim& \mbox{normal}(\mu_{\rm post}^l, \sigma_{\rm post} ),
\end{eqnarray}
where 
\begin{align}
\label{eq:gaussian-gaussian-post}
\mu_{\rm post}^l=\frac{y^l}{1+\sigma^2} \ \ \mbox{ and } \ \
\sigma_{\rm post}= \sqrt{\frac{\sigma^2}{1+\sigma^2}}.    
\end{align}

In the limit of large $S$, the $z$-score of $\theta^l$ within the simulations of $\theta_{\rm post}^{l,s}$ is
\begin{equation}
\label{z_score} z^l = \frac{1}{\sigma_{\rm post}}(\theta^l - \mu^l_{\rm post}).
\end{equation}
The corresponding quantiles are $\Phi(z^l)$, but since we are working here entirely with normal distributions it will be simpler to stick with $z$-scores. 

For prior calibration checking, we must now look at the distribution of these  $z$-scores, averaging $\theta^l$ and $y^l$ over  (\ref{model_1a}) and  (\ref{model_1b}).  The result is that $\theta^l - \mu^l_{\rm post} \sim \mbox{normal}(0, \sigma_{\rm post})$, 
hence the $z$-score (\ref{z_score}) has a unit normal distribution in the limit of a large number of simulation draws $S$.

\subsection{Posterior calibration checking}
\label{sec:posterior_calibration_checking}

\begin{figure}
\includegraphics[width=.9\textwidth]{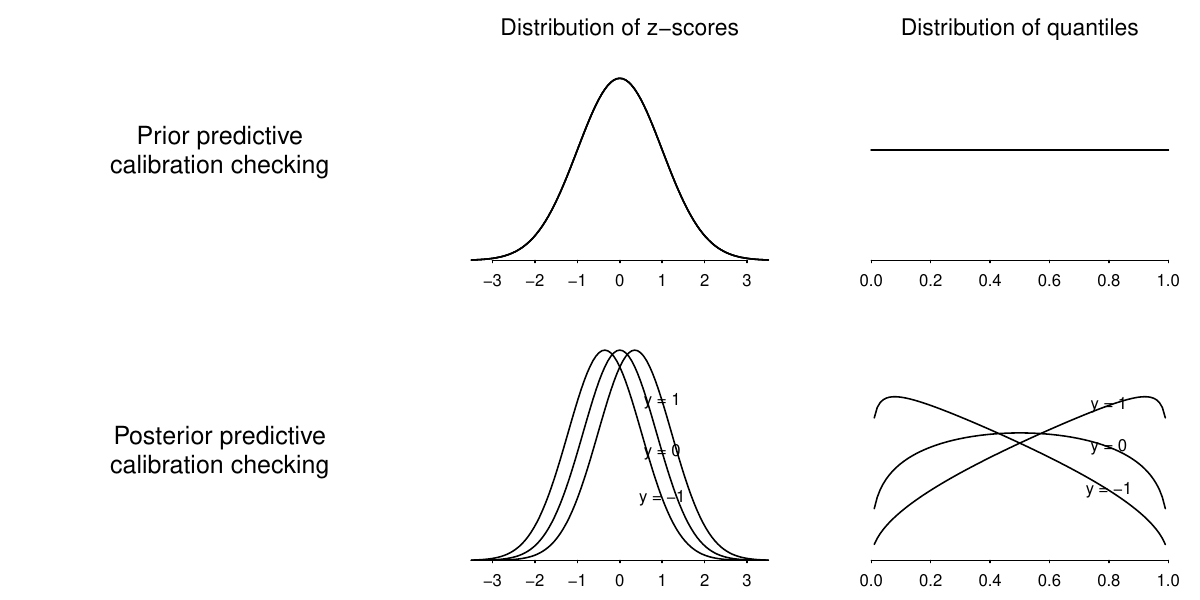}
  \caption{\em Distribution of $z$-scores and quantiles in simulation-based calibration checking when the correct model is being fit for the simple normal example, under two scenarios, both for $\sigma=1$.  {\em Top row:}   Simulating the parameter $\theta^l$ from the prior distribution.  {\em Bottom row:}  Simulating $\theta^l$ from the posterior distribution, in which case the distributions depend on the data, $y$.  Curves show distributions conditional on three possible data values.}\label{quantiles_1}
\end{figure}

Next we look at the properties of calibration checking when starting with draws from the {\em posterior} distribution.  All is the same as before except that the prior (\ref{model_1a}) is replaced with the posterior (\ref{model_1c}),
\begin{equation}
  \label{model_2a} \theta^l \sim \mbox{normal}(\mu_{\rm post}, \sigma_{\rm post} )
\end{equation}
where $\mu_{\rm post}=\frac{y}{1+\sigma^2}$ and $\sigma_{\rm post}=\sqrt{\frac{\sigma^2}{1+\sigma^2}}$ as in \eqref{eq:gaussian-gaussian-post}.
The $z$-score of $\theta^l$ within the simulations $\theta^{l,s}_{\po},s=1,\dots,S$ is still given by (\ref{z_score}); the only difference is that, instead of averaging over  (\ref{model_1a}) and  (\ref{model_1b}), we average over (\ref{model_2a}) and (\ref{model_1b}).  
In this posterior predictive distribution, some algebra leads to
\begin{align}
z^l = \frac{1}{\sigma_{\rm post}}(\theta^l - \mu_{\rm post}^l) \sim 
\mbox{normal}\left(\frac{\sigma}{(1+\sigma^2)^{3/2}}\,y, \frac{\sqrt{\sigma^4+\sigma^2+1}}{1+\sigma^2}\right).
\label{eq:from_posterior}    
\end{align}
For no value of $y$ will this be a unit normal distribution (for $0<\sigma<\infty $), thus we would {\em not} see predictive calibration (a uniform distribution of the $L$ quantiles), even if the computation is exact; see remark below.

\begin{remark}[Derivation of \eqref{eq:from_posterior}]
Let $z_1$ and $z_2$ be independent draws from the unit normal distribution and then do the following steps:
\begin{enumerate}
    \item Draw $\theta^l\sim \mbox{normal}(y/(1+\sigma^2), \sigma / \sqrt{1+\sigma^2})$; equivalently,  $\theta^l=\frac{y}{1+\sigma^2}+\frac{\sigma}{\sqrt{1+\sigma^2}} z_1$
    \item Draw $y^l|  \theta^l = \mbox{normal}(\theta^l, \sigma)$; equivalently, $y^l = \theta^l + \sigma z_2$
    \item Compute $z^l=\frac{1}{\sigma_{\rm post}}(\theta^l - \mu_{\rm post}^l)$; here, $\mu_{\rm post}^l=\frac{y^l}{1+\sigma^2} = \frac{\theta^l+\sigma z_2}{1+\sigma^2}$. Then $\theta^l-\mu_{\rm post}^l = \theta^l \left(1-\frac{1}{1+\sigma^2}\right) - \frac{\sigma}{1+\sigma^2}z_2$. Then substitute $\theta^l=\frac{y}{1+\sigma^2}+\frac{\sigma}{\sqrt{1+\sigma^2}} z_1$
    and use $\frac{1}{\sigma_{\rm post}}=\frac{\sqrt{ (\sigma^2+1)}}{\sigma}$ 
    to rewrite $$z^l=\frac{\sqrt{\sigma^2+1}}{\sigma} \left( \frac{y}{1+\sigma^2} \frac{\sigma^2}{1+\sigma^2} + \frac{\sigma}{\sqrt{1+\sigma^2}}\frac{\sigma^2}{1+\sigma^2} z_1 - \frac{\sigma}{1+\sigma^2}z_2 \right),$$
    which has mean $\frac{\sqrt{\sigma^2+1}}{\sigma} \frac{y}{1+\sigma^2} \frac{\sigma^2}{1+\sigma^2}=\frac{\sigma}{(1+\sigma^2)^{3/2}y}$
    and variance
    $\frac{{\sigma^2+1}}{\sigma^2} \left(
    \frac{\sigma^6}{(1+\sigma^2)^3}
    +
    \frac{\sigma^2}{(1+\sigma^2)^2}
    \right)=\frac{{\sigma^4+\sigma^2+1}}{(1+\sigma^2)^2}$.
\end{enumerate}
\end{remark}

The bottom row of Figure \ref{quantiles_1} shows the distribution of the $z$-scores and quantiles for different values of $y$, for $\sigma=1$.  For comparison, the top row of the figure shows the corresponding curves for the calibrated case of prior predictive checking.

\begin{figure}
\centerline{\includegraphics[width=.8\textwidth]{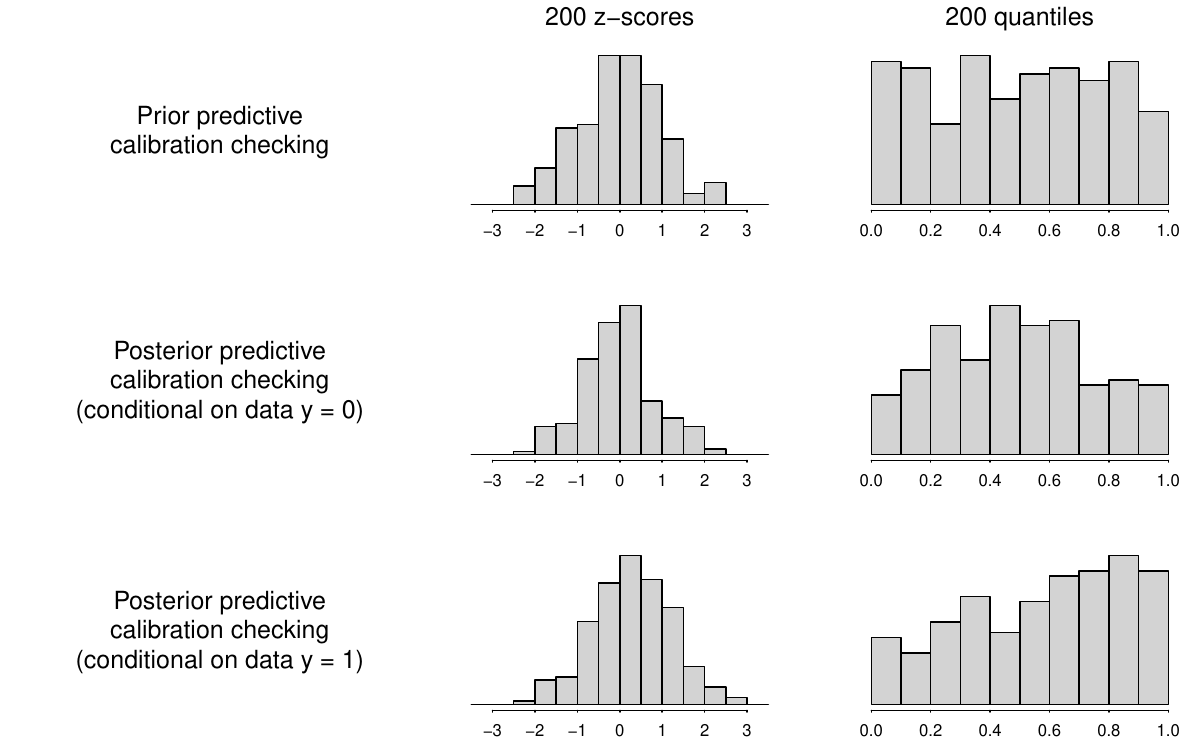}}
  \caption{\em Simulation-based calibration checking using 200 replications of 1000 simulation draws, for each row, with $\sigma=1$.  {\em Top row:}   Simulating the parameter $\theta^l$ from the prior distribution.  {\em Center and bottom rows:}  Simulating $\theta^l$ from the posterior distribution, in which case the distributions depend on the data, $y$.  As predicted by theory, the prior predictive simulations show calibration and the posterior predictive simulations do not, for $\sigma=1$.}\label{quantiles_2}
\end{figure}

Figure \ref{quantiles_2} shows corresponding results from a simulation with $\sigma=1$, $L=200$ and $S=1000$, first drawing $\theta^l$ from the prior, which correctly shows calibration, then drawing from two different posteriors, one conditional on data $y=0$, the other conditional on $y=1$.  The posterior $z$-scores and quantiles do not show calibration, despite that there is no approximation in the simulations.

These $z$-scores \emph{are} a standard normal when $\sigma=0$ and when $\sigma\to\infty$, indicating that posterior calibration checking is valid in the extreme cases in which the prior or the data are uninformative. 

\subsection{Recalibration using a location-scale shift}
The idea of simulation-based recalibration is to use any miscalibration found in the checking to adjust the simulation draws.  Here we use Algorithm~\ref{alg:adjust} in combination with simple location-scale shift as described in Remark~\ref{rmk:zscore}; in Section~\ref{sec:zscore_method} we had proposed a simple version with just rescaling and without shift. 


As before, the first step is to perform calibration checking, in which we obtain $z$-scores $z^l,l=1,\dots,L$, and summarize the miscalibration by $\bar{z}$ and $s_z$, the mean and standard deviation of these $L$ values. The next step is to alter the fitting procedure by dilating the simulations $\theta_{\rm post}^{l,s}$ by the factor $s_z$ and shifting them by the relative value $\bar{z}$.  In this affine transformation, draws from the approximate posterior $\theta^l_{\po}$ are replaced by these adjusted draws:
\begin{equation}
  \label{calibrate}
  \theta_{\rm post}^{{\rm adj}, l,s} = \mu_{\rm post}^l + s_z(\theta_{\rm post}^{l,s} - \mu_{\rm post}^l) + \bar{z} \sigma_{\rm post}^l\,,
\end{equation}
where $\mu_{\rm post}^l$ and $\sigma_{\rm post}^l$ are the mean and standard deviation of the $S$ values, $\theta_{\rm post}^{l,s}$. Under calibration, $\bar{z}$ and $s_z$ should be approximately 0 and 1, respectively, in which case the adjustment should essentially do nothing.  If the fitting procedure is off in its first two moments, this recalibration should correct for that. This is the same adjustment and conditions as in Equations~\eqref{eq:z_score_condition} and \eqref{eq:z_score_adjust_both}, except when draws are from the posterior, rather than the prior. 

If we apply the calibration procedure to the prior predictive distribution, there should be essentially no effect.  More precisely, there will be some random adjustments because of finite number of simulations---Monte Carlo error---but these adjustments should be minor.

\begin{figure}
\centerline{  \includegraphics[width=\textwidth]{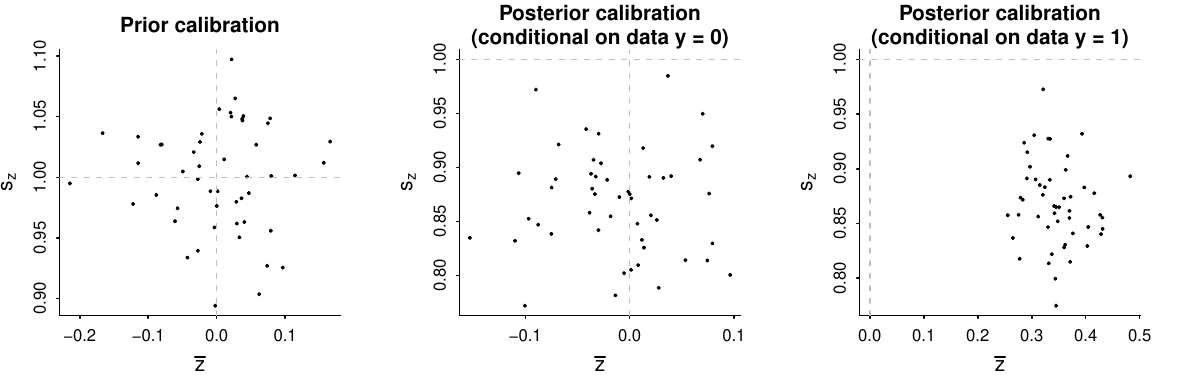}}
  \caption{\em Estimated mean and scale shifts from simulation-based calibration using $L=200$ replications of $S=1000$ simulation draws.  In each scatterplot, the dots correspond to 50 independent simulations of the entire calibration process.  {\em Left:}   Simulating the parameter $\theta^l$ from the prior distribution; adjustments are close to the null adjustment, $(\bar{z}, s_z)=(0,1)$.  {\em Center and right:}  Simulating $\theta^l$ from the posterior distribution, in which case the distributions depend on the data, $y$.  In these cases, the recalibration has a large effect on posterior inferences.}\label{adjusted}
\end{figure}

We check this by performing this recalibration using simulations from model (\ref{model.0}) with $\sigma=1$, using $L=200$ prior simulation draws and $S=1000$ draws from each posterior, replicating the entire procedure 50 times.  For each, we calculate $\bar{z}$ and $s_z$.  The left plot in Figure \ref{adjusted} shows the results; most of the time the mean shift is less than 0.1 and the scale shift is less than 5\%.

\subsection{Evaluating posterior recalibration}

What happens if we try to calibrate based on the posterior quantile?  From a Bayesian standpoint, this is the wrong thing to do, but we can see what happens.  As before, we set $\sigma=1$, $L=100$, $S=1000$ and then loop the entire process 50 times to see what could happen.

The center and right plots in Figure \ref{adjusted} show the results conditional on data $y=0$ and $y=1$, respectively.  To understand what is happening, we consider a typical value for $(\bar{z}, s_z)$ in each case:
\begin{itemize}
\item Given data $y=0$, the correct posterior distribution is $\theta| y\sim\mbox{normal}(0, 0.71)$.  Shifting this by $\bar{z}s_{\rm post}=0$ and scaling it by $s_z=0.87$ yields an adjusted posterior of $\mbox{normal}(0, 0.61)$.
\item Given data $y=1$, the correct posterior distribution is $\theta| y\sim\mbox{normal}(0.5, 0.71)$.  Shifting this by $\bar{z}s_{\rm post}=0.35\cdot 0.71=0.25$ and scaling it by $s_z=0.87$ yields an adjusted posterior of $\mbox{normal}(0.75, 0.61)$.
\end{itemize}
Those simulations show what could happen with $L=200$ in two special cases, $y=0$ and $y=1$.

For general $y$, we can work out analytically the adjustment that would occur in the limit of large $L$ and $S$.  Given $\theta^l$ and $y^l$, the $z$-score of $\theta^l$ in the limit of large $S$ is $z^l=\sqrt{2}(\theta^l - y^l/2)$, from (\ref{model_2a}).  We can figure out the mean and standard deviation of this distribution, averaging over the posterior predictive distribution, in two steps.  First we average over $y^l| \theta^l\sim\mbox{normal}(\theta^l,1)$.  Propagating that uncertainty, $z^l$ has mean $\theta^l/(2\sqrt{2})$ and standard deviation $1/(2\sqrt{2})$.  Next we average over $\theta^l\sim\mbox{normal}(y/2, 1/\sqrt{2})$.  Propagating that uncertainty, $z^l$ ends up with mean $y/(2\sqrt{2})$ and standard deviation $\sqrt{3}/2$.

If we apply posterior recalibration in our problem, it will change the posterior inferences in two ways.  First, there will be much less partial pooling.  Instead of the posterior mean of $\theta$ being $y/2$, it becomes $(\frac{1}{2} + \frac{1}{2\sqrt{2}}\frac{1}{\sqrt{2}})y = \frac{3}{4} y$.  Second, uncertainty will be understated.  Instead of the posterior standard deviation of $\theta$ being $\frac{1}{\sqrt{2}}=0.71$, it becomes $\frac{1}{\sqrt{2}}\frac{\sqrt{3}}{2}=0.61$.

To understand how this happens, start with the actual posterior, $\theta| y\sim\mbox{normal}(y/2, 1/\sqrt{2})$.  In this case, simulated data $y^l$ will be centered around $y/2$, and posterior inferences for $\theta_{\rm post}$ will again be partially-pooled toward zero and will be centered around $y/4$.  The values of $\theta$ drawn from the posterior distribution will be systematically farther from the prior, compared to draws from $\theta_{\rm post}$, and the calibration procedure will then adjust for this by pulling the posterior simulations away from the prior, thus reducing the amount of pooling.  In this case, the recalibrated intervals pool only half as much as the correct posterior intervals.

\subsection{Why posterior calibration can approximately work for hierarchical models}\label{posterior.hier}

Posterior recalibration can work well by making use of internal replication.  Consider a hierarchical model with hyperparameters $\phi$, local parameters $\alpha_1,\dots,\alpha_J$, and local data $y_1,\dots,y_J$.  When drawing from the posterior, we can draw from the $\alpha_j$'s for the $J$ existing groups (in which case the inference for each $\alpha_j$ will be conditional on the observed $y_j$) or for $J$ new groups (in which case the inference for each $\alpha_j$ will be from its prior, conditional on $\phi$).

Now suppose you have enough data so that the hyperparameters $\phi$ are precisely estimated in the posterior, so that $p(\phi|  y_{1:J})$ 
can be approximated by a delta function at some $\hat\phi$.  
In that case, $p(\alpha|   y_{1:J})\approx p(\alpha|  \hat \phi, y_{1:J})$.
Furthermore, posterior draws of $\alpha$ for $J$ new groups are essentially the same as prior draws---for a prior with fixed, correct hyperparameters $\phi\approx \hat\phi$---since $p(\alpha|  y_{1:J})\approx p(\alpha|  \hat\phi, y_{1:J})\approx p(\alpha|  \hat\phi)$. In addition, if the model is correct and $J$ is large enough, then posterior draws for $\alpha$ for the $J$ existing groups will approximate a set of $J$ draws from the prior.  Thus, in this setting, posterior recalibration checking for a hierarchical model should look a lot like prior recalibration checking. 

Our result is distressing---the adjustment from posterior recalibration shifts the posterior distribution and also makes it overconfident!  If we want to make an argument in favor of this procedure, we could say that practitioners often have a desire to perform less partial pooling than is recommended by Bayesian inference, in part because of concern about calibration, that intervals will not have nominal coverage conditional on the true parameter value, rather than nominal coverage averaged over the prior. In straight Bayesian inference, intervals have nominal coverage conditional on the data.  Calibration that creates nominal coverage under the posterior could be thought of as a sort of compromise, a Bayesian analogue to classical coverage, and a step toward some form of generalized Bayesian inference.

We are not fully convinced by this argument, as we went into this problem with the simpler goal of improving  approximations to Bayesian inference.  Still, we have seen examples where going outside the Bayesian framework can systematically improve predictions in a model-open setting \citep{yao2018using,yao2022bayesian}, so we are open to the idea that posterior recalibration could serve some robustness goal.  

\subsection{Prior or posterior recalibration?}
\label{sec:prior_or_posterior}

Our motivation for simulation-based recalibration of inferences is that we often use approximate computational algorithms, and we would like to ensure that the computation can approximately recover parameter values by adjusting for inferential biases in known settings that are relevant to the data at hand.
In general it is not possible to correct an approximate distribution in high dimensions.  The hope with simulation-based recalibration is that it could be possible to attain approximately nominal coverage for scalar summaries, one at a time, without aiming to solve the 
problem of calibrating the joint distribution.

Bayesian inference is automatically calibrated when averaging over the prior distribution.  However, in complicated models, the prior distribution can be broad and include regions of parameter space that are not at all supported by the data.  For any particular dataset, we will not care about the performance of the inference algorithm in these faraway places; rather, we want to focus our checking effort on the region of parameter space that is consistent with the model and data:  the posterior distribution.

However, Bayesian inferences will not be calibrated when averaging over the posterior, as is well known in theory and demonstrated in the present paper.  
The fundamental problem with posterior recalibration---that, from a Bayesian perspective, we would not expect or even want intervals to have nominal coverage averaging over the posterior---is similar to the calibration problems of posterior predictive $p$-values \citep{gelman1996posterior}.  Methods have been proposed to recalibrate predictive $p$-values to be uniformly distributed \citep{robins2000asymptotic,hjort2006post}; it has also been suggested that a uniform distribution of $p$-values is not necessarily desirable for the goal of predictive model evaluation \citep{gelman2013two}.  

It is an open question how best to adjust posterior intervals in a way that incorporates the posterior.
In posterior simulation-based calibration \citep{Sailynoja-Schmitt-Burkner-Vehtari:2025}, SBC is performed to check calibration but where the posterior is used as a new prior. Unlike our discussion of posterior calibration (Section~\ref{sec:posterior_calibration_checking}), where the posterior is used in place of the prior for averaging (Step 1 of Algorithm~\ref{alg:sbcc}), \citet{Sailynoja-Schmitt-Burkner-Vehtari:2025} use the posterior in place of the prior for averaging \emph{and} for posterior inference (Step 4 in Algorithm~\ref{alg:sbcc}). By doing this, they focus calibration checking on a zone of parameter space consistent with the data, and retain the validity of SBC, at the cost of changing the quantity of interest for which they check calibration. \citet{bls2}   use a similar approach of replacing the prior entirely with the posterior in both the averaging and posterior inference steps of SBC, but go one step further to adjust posterior intervals, rather than just check for calibration. Similar to \citet{Sailynoja-Schmitt-Burkner-Vehtari:2025}, \citet{bls} change the quantity of interest, and then use the appropriate scale and shift for this new quantity of interest for adjusting posterior intervals for the original inference problem. 

\section{Discussion}

Bayesian inference is automatically calibrated when averaging over the prior predictive distribution; from that perspective, calibration is a concern only to the extent that the prior and data models might be wrong (that is, robustness of inferences) and if there might be problems with computation.  Posterior recalibration does not play any role in formal Bayesian inference; indeed, with proper and nondegenerate priors, Bayesian inferences cannot be calibrated under the posterior predictive distribution for the same reason that the posterior mean cannot be a classically unbiased estimate.  The contribution of the present paper is to demonstrate and explore posterior miscalibration in a simple example.

That said, there can be reasons for studying posterior recalibration.  From the standpoint of Bayesian workflow, it can make sense to study the performance of approximate computation in the zone of parameter space that is consistent with the data.  More generally, calibration of forecasting and uncertainty estimation is a goal in its own right, not just restricted to Bayesian inference \citep{gneiting2007,cockayne2022testing}.  Finally, posterior recalibration could be a useful tool in hierarchical models, as discussed in Section \ref{posterior.hier} and building off the posterior calibration checking ideas of \cite{Sailynoja-Schmitt-Burkner-Vehtari:2025}, which are extended for recalibration by \cite{bls}.

\subsection{Limitations of proposed calibration methods}

In this work we explored simple methods to recalibrate approximate posteriors that are based on simulation-based calibration checking \citep{Modrak2022SimulationBasedCC,talts2020validating}. Some limitations arise from inherent assumptions around posterior calibration. First, we have assumed that the model is correct, as calibration is defined with respect to data that are generated from parameters drawn from the prior. Second, as calibration is defined on average over parameters drawn from the prior, each individual adjusted posterior may not be correct. 

Other limitations are due to details of our method. For example, it's possible that a universal proportional rescaling around the posterior mean is not the appropriate adjustment: perhaps the correct rescaling varies over data draws, or perhaps the posterior intervals should be widened but by different amounts for different $(1-\alpha)$ intervals. In the absence of additional knowledge, a naive rescaling around the posterior mean seems like a reasonable, if imperfect, choice. Our method also only looks at rescaling, though it can be extended to include shifts or adjustments for higher moments. Lastly, we focus on posterior calibration for one parameter or scalar quantity of interest at a time. We do not try to attain calibration of joint posterior probabilities in higher dimensions.

\subsection{Open questions of proposed calibration methods}
 
Many open questions remain. For example, if we are to find the best rescaling, what is the most sample-efficient way to do so? When should we use either of the two methods proposed in Section \ref{sec:methods} (or another variant entirely), how flexible should a recalibration procedure be, when might recalibration be more vs less effective, and how can we evaluate recalibration procedures more generally?  

In the present paper we have aimed to develop understanding of prior and posterior recalibration focusing on interval coverage of parameters or quantities of interest, one at a time.  Even if our recalibration procedure works perfectly, it will not yield a set of recalibrated joint simulations, a point that also arises in the work of \cite{fearnhead2012} and \cite{li2017}.


\if 1=0
\paragraph{Unsorted for now}
Prangle 2014

https://arxiv.org/pdf/1507.05179.pdf

https://arxiv.org/pdf/1711.11057.pdf

http://www.stat.columbia.edu/\~gelman/research/published/mi\_theory9.pdf

https://arxiv.org/pdf/1810.06433.pdf

http://proceedings.mlr.press/v97/xing19a/xing19a.pdf

\fi
\if 1=0
\section{What should we be calibrated with respect to?}
By this question, we mean, from what distribution should we draw $\tilde \theta$,
in order to learn the adjustment to posteriors, in order to calibrate them? 

A natural choice is the prior. 
If $M=M'$, then exact Bayesian posteriors are calibrated with respect to the prior. 
Hopefully, this prior is chosen well. 

However, despite these reasons for calibrating with respect to the prior, 
it can be hard to choose a prior well, so that reasonable choices of prior may put too much mass in areas of $\tilde\theta$ that we are uninterested in. Instead, one might argue that it's better to draw 
$\tilde\theta|  y_D$ from the posterior, conditioned on the observed data, instead. However, the posterior on $\theta$ given the observed data can also put too little weight on the true $\theta$: consider a simple example in which $y\sim\mbox{normal}(\theta, 1),\, \theta\sim\mbox{normal}(0,1)$. If $\theta$ is truly 0, it is plausible to draw $y=2.1$, but the posterior is then $\theta|  y\sim \mbox{normal}(1.05, 0.71)$, which puts little weight in the neighborhood of the true $\theta=0$.

At the same time, there are settings in which calibration with respect to the posterior is appropriate, asymptotically \tc{todo}.

\section{Theoretical results}
\subsection{Theoretical results when drawing from the correct distribution}
So far we have presented a method and some examples, along with some intuition.

What can be proved?

If $M^{\prime}\equiv M$ and we are drawing from the {\em prior}, we should have exact calibration (in expectation) in our original inferences, and the calibration adjustment in this paper will only make things worse, but hopefully not much worse. 
Since posterior intervals are calibrated by construction, 
then asymptotically, with more draws from the prior, both methods from Section~\ref{sec:methods} would do nothing.

\begin{prop}
As sample sizes $n,N\to\infty$, the adjustment learned in Algorithm~\ref{alg:adjust}
approaches the identity. \tc{todo: make this rigorous}
\end{prop}

If $M^{\prime}\equiv M$ and we are drawing from the {\em posterior}, we will {\em not} in general have calibration, as can be seen from a simple example, $y\sim\mbox{normal}(\theta, 1),\, \theta\sim\mbox{normal}(0,1)$.  We can talk about how bad this adjustment can get. \tc{I think you'd actually be fine here.}

Asymptotically the miscalibration should go away. \tc{I don't understand this}

\subsection{Theoretical results when drawing from an approximate distribution}

To be able to use a method with confidence, it is necessary to know where it fails.

\subsection{Checking coverage for a vector of parameters}

For a multilevel model, there are conditions in which the calibration will be good even for small samples, if the number of groups is large.  The idea is that if we are calibrating a quantity that has internal replication, then we are essentially averaging over its prior (conditional on hyperparameters) so we get the appropriate coverage.

Bounds on coverage for each element of the vector \dots

Average coverage for all elements \dots
\fi 
\bibliography{bib}

\end{document}